# Properties of Chromatin in Human Cells as Characteristics of the State of Human Organism: A Review

Shckorbatov Y.G.

V.N. Karazin Kharkiv National University, Kharkiv, 61022, Ukraine
yuriy.shckorbatov@gmail.com

*Abstract:* The state of chromatin in individual cells is directly related to the state of the whole organism and may be used for assessment of the state of the whole organism. Many hereditary diseases are connected with chromatin rearrangements. Chromatin distribution in cell nucleus is an important characteristic which may be applied for determination of disease gravity. The method of determination of the quantity of heterochromatin granules in buccal epithelium cells may be applied in the assessment of the state of human organism during sportive trainings and medical treatment.

**Introduction**

In his famous work "Cell Pathology" Rudolf Virchow outlined that the pathology of organism begins as pathology of cells [1]. One of important characteristics of the cell is the state of chromatin in cell nucleus. Chromatin is a complex of DNA with nuclear proteins. It may be divided in two forms: a dense, low-activity state (heterochromatin) and a diffused, active state (euchromatin) [2]. The increase of portion of heterochromatin in cell nucleus is a sign of decrease of processes of RNA synthesis, or in other words, it is a sign of partial inactivation of the cell nucleus. Below some results demonstrating the role of chromatin rearrangements in pathology and sport medicine are presented.

**Discussion**

Chromatin remodelling is often connected with hereditary diseases. The role of mutations of proteins associated with the nuclear membrane and the corresponding violations of the nuclear membrane and chromatin structure are described for many hereditary diseases. Among them groups of diseases related to striated muscles, adipose tissue, peripheral nerves, progeria phenotype [3]. Some of the diseases are connected with mutations of A-type and B-type lamins (so-called laminopathies) are connected to changes in heterochromatin layer adjacent to the nuclear membrane [4][5]. The role of different nuclear proteins in peripheral heterochromatin formation in granulocytes is in details described in [6]. The significance of the nuclear actin structures in regulation of cell nucleus morphology, gene expression and cell answer to disease and environmental agents is discussed in [7].
The well-studied autoimmune disease systemic lupus erythematosus (SLE) which is characterized by producing of autoantibodies to the cell nucleus is also associated with changes in the structure of the nuclear membrane and chromatin. Patients with SLE have neutrophils with dumbbell-shaped or ovoid nuclei that resembled the Pelger–Huet anomaly. The Pelger–Huet anomaly is usually autosomal dominant trait. This condition is characterized by granulocytes that are either bilobed or completely unsegmented [8].The changes in nuclear morphology related to SLE are connected with changes in splicing of mRNA of the nuclear membrane protein - lamin B receptor (LBR) [9].
Heart failure involved decreased stability of chromatin interactions around disease-causing genes. In addition, pressure overload or CTCF depletion remodelled long-range interactions of cardiac enhancers, resulting in a significant decrease in local chromatin interactions around these functional elements. It was demonstrated that global structural remodelling of chromatin underpins heart failure [10].

A marker for assessment of cell nucleus morphology changes, termed Nucleotyping was used in [11], based on automatic assessment of disordered chromatin organization. This marker stratifies on regions defined by the distance to the nuclear periphery. The chromatin organisation in cell nuclei was automatically analysed by examining the spatial variations in DNA density. The pattern of spatial distribution of DNA in the nucleus was assessed for each prostate cancer patient. By this pattern all the patients after radical prostatectomy were distributed in two groups and for each group the risk or recurrence was calculated. Nucleotyping predicted recurrence with a hazard ratio (HR) of 3.3 (95% confidence interval (CI), 2.1–5.1) [11].

In a series of works of scientific group in Kharkiv National University were assessed the chromatin changes in human buccal epithelium cells, connected with changes of the state of organism due to ageing, sportive charges, visual charges, and disease. The cells of the human buccal epithelium are good object for studying the processes of euchromatin ↔ heterochromatin transitions, since they have a large flat shaped nucleus with well-structured chromatin. It was proposed to assess the state of chromatin in cell nucleus by the only parameter, the heterochromatin granule quantity (HGQ) after dyeing the cells with orcein [12]. The heterochromatin granules quantity in the nucleus is evaluated in 30-50 nuclei and the mean value of the HGQ and the standard error of the mean HGQ are determined. Normally, the standard error for each experiment not exceeds 5% of the mean HGQ value.

The age related chromatin condensation in cells of healthy donors of cells was demonstrated [13]. A Patent of Ukraine was obtained for "Method for determining of fatigue in humans" in which the method of determination of the HGQ degree was applied for assessment of the effect of physical loadings [14]. The chromatin condensation in exfoliated buccal epithelium cells was revealed after sportive trainings. Cells were collected during training crosses from students and lecturers of the Kharkiv Institute of Air Forces of Ukraine. The cross lasted for 8-10 h. The energy waste for one person was in different cases approximately 1 or 2 MJ. These physical loadings resulted in the significant increase of HGQ. After the period of repose (16 h) the chromatin condensation in most cases reversed to the initial (control) level [15].

The method of HGQ was applied for assessment of the state of organism of cancer patients. As it is known homocysteine plays an important role in tumor genesis, it inhibits angiogenesis and tumor invasion. Among patients – women with breast neoplasms – breast cancer and fibroadenoma patients, the homocysteine level was determined in blood plasma. An inverse correlation was found between heterochromatin granule level in buccal epithelium nuclei, and homocysteine level in blood plasma [16].

HGQ was estimated in the investigation carried out on a stationary bicycle stand trainer in a group of students of Kharkiv Medical University (13 men) of age 19-23. Sport exercises on stationary bicycle stand trainer VE02 (Kiev) at power charge 400 W for 1 min. In the most of donors HGQ increased but no increase of HGQ was observed in two donors, proved to be sportsmen [17]. The method of HGQ assessment was used in the study of the influence of working with the text on the screen of the computer monitor on the functional state of an operator. It was shown that the HGQ increases after visual work with text with low readability. On the contrary, after the computer game "Kosynka" the HGQ is decreasing, which, in authors' opinion, is due to the positive emotional background of this kind of activity. The obtained results allow to state that the text presented on the electronic information carrier has a greater negative impact on the state of human organism than the text printed on paper, causing the development of visual fatigue in 70% of the subjects. Using the HGQ opens the new opportunities for understanding the mechanisms of adaptation to different types of visual load [18]. The HGQ was assessed in cells of human buccal epithelium among trained and untrained students during volleyball and tennis training sessions: before, in the process and after training session. HGQ increased during training sessions among both volleyball and tennis players. This parameter increased with time of performance of the training session. During the training it was observed the uniform increase in the HGQ during training and after training for a group of untrained students, and during the first steps of training among the athletes was registered a small increase in the HGQ, compared to a group of untrained students. This dynamics of HGQ changing in different groups of players suggests that the organism of athletes answer to training loadings more slowly than it is observed among group of untrained students. At the end of the training, the HGQ score has the same average level in both groups [19]. The measurement of HGQ during military training shows that the growth of HGQ in buccal epithelium cells under the influence of psychophysical stress. The stress can be estimated at the level of "high" without approaching the limits of endurance. The research during a raid in extreme conditions showed that the limit of endurance (up to 5 mJ) the HGQ increased to the limit values at 22-23, which in some cases was characterized by a partial loss of coordination and orientation in space, inadequate response to commands and practical inability to use a systematic approach to make a decision [20].

The method of HGQ determination in exfoliated cells was also applied in observations among patients with neurological disorders. It was shown that among patients with diffused sclerosis the HGQ level is significantly elevated in comparison with control group [21]. The analysis of the data obtained among the patients with Wilson's disease (hepatolenticular degeneration) revealed that HGQ in buccal epithelial cells was significantly lower than in control group [22].

The method of chromatin state assessment was applied in investigation of the state of buccal epithelium cells in different groups of cigarette and hookah smokers in comparison with control group of non-smokers. It was shown the increase of HGQ (chromatin condensation) in cells of buccal epithelium among hookah smokers [23].

Summing up, the increase of the degree of heterochromatinisation of chromatin in buccal epithelium cell nuclei indicates cell response to stress at the cellular level. The degree of heterochromatisation increases in unspecific response to different unfavorable influences on human organism. In our opinion, mentioned above results are in a good agreement with well-known views of Rudolf Virchow about cellular mechanisms of pathologies.

**Conclusion**

Different experimental approaches demonstrate the significant role of chromatin rearrangements in cells corresponding to diseases and physiological stress at organism level. Assessment of changes in chromatin may give clues to more propound understanding of the mechanisms of pathology. On the other side assessment of chromatin rearrangements may be applied for diagnostics of the gravity of disease and also may be used in sportive and occupational medicine.

*Funding:* This work was supported by the Ministry of Education and Science of Ukraine (Grants 0115U000487; 0117U004831).

**References**

[1] Virchow R (1858) Die Cellularpathologie in ihrer Begründung auf physiologische und pathologische Gewebelehre. Verlag von August Hirschwald, Berlin.
[2] Krebs J, Kilpatrick S Goldstein E (2014) Lewin's Genes XI, Jones & Bartlett Learning Burlington, MA.
[3] Worman HJ, Östlund C, Wang Y (2010) Diseases of the Nuclear Envelope. Cold Spring Harbor Perspectives in Biology 2(2): a000760. doi:10.1101/cshperspect.a000760
[4] Davidson PM, Lammerding J (2014) Broken nuclei – lamins, nuclear mechanics, and disease. Trends Cell Biol 24(4): 247-256. doi: 10.1016/j.tcb.2013.11.004.
[5] Schreiber KH, Kennedy BK (2013) When lamins go bad: nuclear structure and disease. Cell 152(6): 1365-1375. doi: 10.1016/j.cell.2013.02.015.

[6] Olins AL, Zwerger M, Herrmann H, et al. The human granulocyte nucleus: unusual nuclear envelope and heterochromatin composition. European journal of cell biology. 2008;87(5):279-290. doi:10.1016/j.ejcb.2008.02.007;

[7] Serebryannyy LA, Yuen M, Parilla M, Cooper ST, de Lanerolle P (2016) The effects of disease models of nuclear actin polymerization on the nucleus. Front Physiol 7: 454. eCollection 2016.

[8] Shah SS, Parikh RS, Vaswani LP, Divkar R (2016) Familial Pelger-Huet anomaly. Indian J Hematol Blood Transfus 32(Suppl 1): 347-350. doi: 10.1007/s12288-015-0508-3.

[9] Singh N, Johnstone DB, Martin KA, Tempera I, Kaplan MJ, Denny MF. Alterations in nuclear structure promote lupus autoimmunity in a mouse model. Dis Model Mech. 2016 Aug 1;9(8):885-897. doi: 10.1242/dmm.024851. Epub 2016 Jun 9.

[10] Rosa-Garrido M, Chapski DJ, Schmitt AD, et al. (2017) High-resolution mapping of chromatin conformation in cardiac myocytes reveals structural remodeling of the epigenome in heart failure. Circulation 136(17): 1613-1625. doi: 10.1161/CIRCULATIONAHA.117.029430.

[11] Hveem TS, Kleppe A, Vlatkovic L, et al. (2016) Chromatin changes predict recurrence after radical prostatectomy. Br J Cancer 114(11): 1243-1250. doi: 10.1038/bjc.2016.96.

[12] Shckorbatov YG (1999) He-Ne laser light induced changes in the state of the chromatin in human cells, Naturwissenschaften 86: 450-453.

[13] Shckorbatov YG (2001) Age-related changes in the state of chromatin in human buccal epithelium cells. Abstracts of the 17th World Congress of the International Association of Gerontology. Vancouver, Canada, July 1-6, 2001, Gerontology 47(Suppl. 1): 224-225.

[14] Shckorbatov YG, Shakhbazov VG, Sutyushev TA, Grigorieva NM (2001) Method for determination of tiredness of human person, Ukraine patent number UA 37774 A. Official Bulletin "Promyslova Vlasnist" No 4 (in Ukrainian). http://uapatents.com/3-37774-sposib-viznachennya-vtomlenosti-lyudini.html

[15] Shckorbatov YG, Zhuravleva LA, Navrotskaya VV, et al. (2005) Chromatin structure and the state of human organism, Cell Biol Internat 29: 77-81.

[16] Chekhun VF, Lozovskaya YV, Sevko AL, et al. (2007) Micronuclei and chromatin structure patterns in buccal epithelium cells and relationships with plasma homocysteine levels in female patients with breast tumors. Oncologia (Oncology) 9(4): 311-314 (in Ukrainian).

[17] Shckorbatov Y, Samokhvalov V, Bevziuk D, Kovaliov M (2009) Changes in chromatin state in donors subjected to physical stress. arXiv:0902.0089.

[18] Magda IY, Kaplin IV, Kochina ML, Shckorbatov YG (2011) Evaluation of the functional state of users of information technologies on quantity of granules of heterochromatin in cells of buccal epithelium, Modern Trends in Biological Physics and Chemistry (BPPC – 2011). Materials of VII International Science-Technical Conference. Sevastopol: 112-114 (in Russian).

[19] Magda IY, Koliy SN, Burko VL, Temchenko VO, Shckorbatov YG (2014) Condensation of chromatin in human buccal epithelium cells after training sessions. Biohelikon: Cell Biology 2: a16. http://doi.org/10.5281/zenodo.888125

[20] Penkov VI, Sutiushev TA, Shckorbatov YG (2014) Personnel loading research when overcoming obstacle course, Honor and law. Scientific journal of the Academy of Internal Troops of Ukraine. 2: 36-40 (in Ukrainian).

[21] Voloshina NP, Shckorbatov YG, Gaponov IK (2010) The state of chromatin in the nuclei of buccal epithelium cells as a marker of multiple sclerosis, Ukrainian Neurological Journal 2 47-52 (in Russian).

[22] Voloshina NP, Shckorbatov YG, Voloshin-Gaponov IK, et al. (2014) Links of pathogenesis of development of neurodegenerative process and its clinical manifestation in patients with hepatocerebral dystrophy, Ukrainian Medical Journal (UMJ) 104: 114-118 (in Ukrainian).

[23] Volkova O, Ryabokon E, Magda I, Shckorbatov Y (2017) Impact of smoking habits on the state of chromatin and morphology of buccal epithelial cells among medical students, Georgian Medical News 262: 111-115.